\documentstyle[prl,aps,floats]{revtex}
\begin{document}
\twocolumn[
\begin{centering}
{\large \bf Low-energy elementary excitations of a trapped
Bose-condensed gas}
\vspace{1.5ex}\\
P. \"Ohberg$^{1}$, E.L. Surkov$^{2,3}$, I. Tittonen$^{4}$,
S. Stenholm$^{1}$, M. Wilkens$^{4}$, and G. V.
Shlyapnikov$^{2,3}$
\vspace{1ex}\\
{\em (1)} {\it Helsinki Institute of Physics,
P. O. Box 9, FIN-00014 University of Helsinki, Finland}\\
{\em (2)} {\it Russian Research Center Kurchatov Institute,
Kurchatov Square,
 123182 Moscow, Russia}\\
{\em (3)} {\it FOM Institute for Atomic and Molecular Physics,
Kruislaan 407,
1098 SJ Amsterdam, The Netherlands}\\
{\em (4)} {\it Fakult\"at f\"ur Physik, Universit\"at Konstanz,
D-78434 Konstanz, Germany}
\vspace{0.5ex}\\
\end{centering}
\begin{quote}

{\small We develop the method of finding analytical solutions of
the
Bogolyubov-De Gennes equations for the excitations of a Bose
condensate
in the Thomas-Fermi regime in harmonic traps of any asymmetry
and introduce a classification of eigenstates.
In the case of cylindrical symmetry we emphasize the presence of
an
accidental degeneracy in the excitation spectrum at certain
values of
the projection of orbital angular momentum on the symmetry axis
and
discuss possible consequences of the degeneracy in the context of
new signatures of Bose-Einstein condensation.

}
\end{quote}
]

\vspace{4mm}

\narrowtext

The recent realization of Bose-Einstein condensation (BEC) in
trapped alkali atom gases \cite{Cor95,Hul95,Ket95}, followed by
the
second generation of experiments
\cite{jila,Mew96,And96,mit,Jin96},
has opened the possibility of investigating macroscopic quantum
phenomena
in these systems.
For understanding the macroscopic quantum behavior of a trapped
Bose-condensed
gas especially important is the character of elementary
excitations of the
trapped condensate, which to a large extent is predetermined by
the
interaction between atoms.
In dilute gases the interaction is primarily binary and is
characterized by a single parameter, $a$, the $s$-wave scattering
length.
This allows one to develop a transparent theory which can be
tested
experimentally.

At present theoretical investigations of elementary excitations
of trapped
Bose condensates include analytical solutions for the spectrum of
low-energy
excitations
in spherically symmetric harmonic traps in the Thomas-Fermi
regime \cite{Stringari} and numerical analysis of the
eigenfunctions and
eigenenergies of the excitations in the traps of spherical and
cylindrical
symmetry \cite{Burnett,Singh,Jav,Zoller,You}. In the latter case
the
eigenfrequencies of the lowest excitations, as those measured in
the
JILA \cite{jila,Jin96} and MIT \cite{mit} experiments, have also
been found
analytically \cite{Stringari,Castin96,Kagan97,Zoller}.

Most interesting are the low-energy excitations, i.e., the
excitations with
energies much smaller than the chemical potential (mean field
interaction
between particles), as they are essentially of collective
character.
Previous studies revealed that the eigenfrequencies of condensate
oscillations are strongly different from those of a collisionless
thermal
gas \cite{jila,mit,Jin96}, but are rather close to the
frequencies of
a thermal gas in the hydrodynamic regime \cite{Kagan97,Griffin}.
In this paper we develop the method of finding analytical
solutions of the
Bogolyubov-De Gennes equations
for the spectrum and wavefunctions of the condensate excitations
in the
Thomas-Fermi regime in harmonic
traps of any type of asymmetry and introduce a classification of
eigenstates.
We analyse the structure of the excitation spectrum
in the case of cylindrical symmetry and find an
accidental degeneracy at certain values of the projection of
orbital angular momentum on the symmetry axis.
We address the question of how the accidental degeneracy can
manifest
itself, providing us with a clear distinction between the
condensate
oscillations and the oscillations of a classical gas in the
hydrodynamic
regime.

We consider a Bose-condensed gas in an external harmonic
potential
$V({\bf r})=M\sum_i\omega_i^2r_i^2/2$ with frequencies $\omega_i$
and
assume a pair potential of the atom-atom interaction
 of the form $U({\bf R})=\tilde U\delta({\bf R})$,
where $\tilde U=4\pi\hbar^2a/M$, $a$ is the scattering length and
$M$
the atom mass. Then the grand canonical Hamiltonian of the system
is written as
\begin{equation}    \label{H}
\!\hat H\!\!=\!\!\!\int\!\!d{\bf r}\hat\Psi^{\!\dagger}\!({\bf
r})\!
\!\left[\!-\frac{\hbar^2}{2M}\Delta\!+\!V({\bf
r})\!-\!\mu\!+\!{1\over 2}
\tilde U\hat\Psi^{\!\dagger}\!({\bf r})
\hat\Psi({\bf r})\right]\!\!\hat\Psi({\bf r}),\!
\end{equation}
where $\hat\Psi({\bf r})$ is the field operator of atoms, and
$\mu$ the
chemical potential.
The field operator can be represented as a sum of the
above-condensate
part and the condensate wavefunction
$\Psi_0=\langle\hat\Psi\rangle$,
which is a $c$-number: $\hat\Psi=\hat\Psi^{\prime}+\Psi_0$ (see
\cite{LL}).
Assuming that the condensate density greatly exceeds
the density of above-condensate particles we omit the terms
proportional to $\hat\Psi^{\prime 3}$ and $\hat\Psi^{\prime 4}$
in
Eq.(\ref{H}) and write the Hamiltonian in the form
\begin{eqnarray}
&&\hat H\!\!=\!\!\hat H_{\!0}\!\!+\!\!\!\int\!\!\!d{\bf r}
\!\Big\{\!\hat\Psi^{\prime\dagger}
({\bf r})\!\left[\!-\frac{\hbar^2}{2M}\Delta\!+\!V({\bf
r})\!-\!\mu\right]\!
\!\hat\Psi^{\prime}\!\!\!+\!\frac{1}{2}\tilde U\!\Big[
4|\Psi_{0}({\bf r})|^2
\!\!\times \nonumber \\
&&\hat\Psi'^{\dagger}({\bf r})\hat \Psi'({\bf r})\!+\!\Psi_0^2
\hat \Psi'^{\dagger}({\bf r})\hat\Psi'^{\dagger}({\bf r})\!+
\!\Psi_0^{* 2}\hat \Psi'({\bf r})\hat\Psi'({\bf r})
\Big]\Big\} ;    \label{Hprime}\\
&&\hat H_{\!0}\!\!=\!\!\!\int\!\!d{\bf r}\Psi_{0}^{*}({\bf r})
\!\left[\!-\!\frac{\hbar^2}{2M}\!\Delta\!+\!V({\bf r})\!-\!\mu
\!+\!\frac{1}{2}\tilde U|\Psi_{0}({\bf
r})|^2\!\right]\!\!\Psi_0({\bf r}).
\!\!  \label{H0}
\end{eqnarray}
The Gross-Pitaevskii equation for $\Psi_0$ normalized by the
condition
$\int |\Psi_0({\bf r})|^2d{\bf r}=N_0$ ($N_0$ is the number of
particles in the condensate) follows directly from $\hat{H_0}$ in
Eq.(\ref{H0})
\begin{equation}         \label{GGP}
\left( -{{\hbar^2 }\over {2 M}}\Delta+V({\bf r})+
\tilde U |\Psi_{0}|^{2}\right) \Psi_{0}=\mu \Psi_{0},
\end{equation}
Owing to Eq.(\ref{GGP}) the part of the Hamiltonian, linear
in $\hat\Psi^{\prime}$ (and not included in Eq.(\ref{Hprime})),
becomes
equal to zero.
The Hamiltonian (\ref{Hprime}) is bilinear in the operators
$\hat\Psi^{\prime}$, $\hat\Psi^{\prime\dagger}$ and can be
reduced to a
diagonal form
\begin{equation}          \label{Hdiag}
\hat H=H_{0}+\sum_{\nu}E_{\nu}\hat b^{\dagger}_{\nu}\hat b_{\nu}
\end{equation}
by using the Bogolyubov transformation generalized to
an inhomogeneous case:
$\hat\Psi^{\prime}({\bf r})\!=\!\sum_{\nu}[(u_{\nu}({\bf r})\hat
b_{\nu}
\!-\!v_{\nu}^{*}({\bf r})\hat b_{\nu}^{\dagger}]$ \cite{Gennes}.
Here $\hat b_{\nu}$, $\hat b_{\nu}^{\dagger}$ are creation and
annihilation operators of elementary excitations. The Hamiltonian
takes
the form (\ref{Hdiag}) if the
functions $u_{\nu}$, $v_\nu$ satisfy the equations
\begin{eqnarray}
\left(\!-{{\hbar^{2}}\over {2 M}}\Delta\!+\!V({\bf
r})\right)\!u_{\nu}\!+\!
\tilde U |\Psi_{0}|^{2}
(2u_{\nu}\!-\!v_{\nu})&=&(\mu\!+\!E_{\nu})u_{\nu}
\label{u}\\
\left(\!-{{\hbar^{2}}\over {2 M}}\Delta\!+\!V({\bf
r})\right)\!v_{\nu}\!+\!
\tilde U |\Psi_{0}|^{2}
(2v_{\nu}\!-\!u_{\nu})&=&(\mu\!-\!E_{\nu})v_{\nu}
\label{v}
\end{eqnarray}
($\Psi_0$ is taken real), and are normalized by the condition
\begin{equation}           \label{norm}
\int d{\bf r} (u_{\nu} u^{*}_{\nu'}-v_{\nu}
v^{*}_{\nu'})=\delta_{\nu
\nu'}.
\end{equation}
The Hamiltonian (\ref{Hdiag}) does not contain the recently
discussed term
originating from the presence of the ``momentum'' operator of the
condensate \cite{phase}, since this term does not affect the
elementary
excitations.

Eqs.~(\ref{u}), (\ref{v}) and (\ref{GGP}) represent a complete
set of
equations for finding the wavefunctions $u_{\nu},v_{\nu}$ and
energies $E_{\nu}$ of the excitations.
We will discuss the case of repulsive ($a\!\!>\!\!0$)
interparticle
interaction in the Thomas-Fermi regime
($\mu\!\approx\!n_{0m}\tilde U\!\gg\!\hbar\omega_i$,
$n_{0m}$ is the maximum condensate density), where
the presence of a small parameter
\begin{equation}             \label{eta}
\zeta=\hbar\overline{\omega}/2\mu\ll 1,
\end{equation}
($\overline{\omega}=\prod_i\omega_i^{1/3}$) allows us to
simplify the equations for the elementary excitations.
First, we write Eqs.~(\ref{u}), (\ref{v}) and (\ref{GGP}) in
terms
of dimensionless eigenenergies
$\varepsilon_{\nu}\!=\!E_{\nu}/\hbar\overline{\omega}$
and coordinates $y_i\!=\!r_i/l_i$, where
$l_i\!=\!(2\mu/m\omega_i^2)^{1/2}$ is the characteristic size of
the
condensate in the $i$-th direction:
\begin{eqnarray}
-\zeta^{2} \tilde\Delta
u_{\nu}+y^{2}u_{\nu}+(2u_{\nu}-v_{\nu})\bar
n_{0}&=&(1+2\zeta\varepsilon_{\nu})u_{\nu} \label{udim} \\
-\zeta^{2} \tilde\Delta
v_{\nu}+y^{2}v_{\nu}+(2v_{\nu}-u_{\nu})\bar
n_{0}&=&(1-2\zeta\varepsilon_{\nu})v_{\nu} \label{vdim} \\
-\zeta^{2} \tilde\Delta \Psi_{0}+y^{2}\Psi_{0}+\bar n_{0}\Psi_{0}
&=&
\Psi_{0} \label{GGPdim}.
\end{eqnarray}
Here 
$\tilde\Delta\!=\!\sum_i(\omega_i/\overline{\omega})^2\partial^2/\partial
y_i^2$,
and $y^2\!=\!\sum_iy_i^2$. With the dimensionless condensate
density
$\bar n_{0}\!=\!|\Psi_0({\bf r})|^2/n_{0m}$ from
Eq.(\ref{GGPdim}),
Eqs.~(\ref{udim}) and (\ref{vdim})
are reduced to the fourth-order differential equations for the
functions $f_{\nu\pm}\!=\!u_{\nu}\!\pm v_{\nu}$:
\widetext
\begin{eqnarray}
&&(1\!-\!y^2)\!\left\{\!-\tilde\Delta
f_{+}\!+\!f_{+}\frac{\tilde\Delta\Psi_0}{\Psi_0}\right\}\!+\!\frac{\zeta^2}
{2}\!\left[\!\tilde\Delta^2 f_{+}\!
-\!3\frac{\tilde\Delta\Psi_0}{\Psi_0}
\tilde\Delta f_{+}\!-\!\tilde\Delta
\left(f_{+}\frac{\tilde\Delta\Psi_0}{\Psi_0}\right)+\!3\!\left(
\frac{\tilde\Delta\Psi_0}{\Psi_0}\right)^2\!\!\!f_{+}\!\right]\!\!
=\!2\varepsilon^2\!f_{+}, \!\!     \label{f+} \\
&&\left\{\!-\!\tilde\Delta
(\!1\!\!-\!y^2)f_{-}\!+\!(\!1\!\!-\!y^2)f_{-}
\frac{\tilde\Delta\Psi_0}
{\Psi_0}\!\right\}\!\!+\!\frac{\zeta^2}{2}\!\left[\!
\tilde\Delta^2\!f_{-}\!\!
\!-\!\frac{\tilde\Delta\Psi_0}{\Psi_0}\tilde\Delta f_{-}\!\!
-\!3\tilde\Delta\!\left(\!f_{-}\frac{\tilde\Delta\Psi_0}{\Psi_0}
\!\right)+\!3\!\left(
\!\frac{\tilde\Delta\Psi_0}{\Psi_0}\!\right)^{\!2}\!\!\!f_{-}\!\right]
\!\!=\!2\varepsilon^2\!f_{-}. \!\!    \label{f-}
\end{eqnarray}
\narrowtext
\hspace{-4mm}Here we have omitted the index $\nu$ and written the
terms
proportional to $\zeta^2$ separately.

The low-energy excitations ($E\ll\!\mu$ or
$\varepsilon\zeta\!\ll\!1$)
are primarily localized inside the condensate spatial region.
At characteristic distances from the condensate boundary
\begin{equation}
\delta y\gg {\rm
max}[\varepsilon\zeta,(\zeta/\varepsilon)^{1/2}],
\end{equation}
we can omit all terms proportional to $\zeta^2$ in
Eqs.~(\ref{f+}), (\ref{f-})
and use the Thomas-Fermi approximation for the
condensate wavefunction (see \cite{Silvera,Huse}):
\begin{equation}       \label{Psi0}
\Psi_{0}=\sqrt{n_{0m}(1-y^{2})},\,\,\,y\leq 1,
\end{equation}
following from Eq.(\ref{GGP}) in which the kinetic energy term
$\zeta^2\tilde\Delta\Psi_0$ is neglected.
Then, using the substitution
$f_{\pm}(y)=C_{\pm}(1-y^{2})^{\pm {1\over 2}}W(y)$,
we obtain the equation
\begin{equation}    \label{W}
\hat G W+2\varepsilon^{2}W=0,
\end{equation}
where the operator $\hat G$ is given by
\begin{equation}   \label{G}
\hat G=(1-y^{2})\tilde \Delta-2\sum_i y_i (\omega_i/{\bar
\omega})^2
{\partial}/{\partial y_i}.
\end{equation}
The relation between the normalization coefficients $C_{+}$ and
$C_{-}$
follows from Eqs.~(\ref{udim}), (\ref{vdim}), (\ref{W}) and
(\ref{G}):
\begin{equation}           \label{C}
C_{-}=\varepsilon\zeta C_{+}.
\end{equation}

The solution (\ref{Psi0}) can be used in Eqs.~(\ref{udim}) and
(\ref{vdim})
from the very beginning for finding
the wavefunctions and spectrum of elementary excitations with
energies
$E_{\nu}\gg\hbar\omega_i$. However, for the excitations with
energies
comparable to the trap frequencies this
would lead to an incorrect result. Moreover, such a
procedure makes Eqs.~(\ref{udim}) and (\ref{vdim}) incompatible
with each
other. The physical reason is that the
wavefunctions of such excitations vary over a distance comparable
with
the
size of the condensate. Hence,
the kinetic energy of the condensate, omitted in the
derivation of Eq.(\ref{Psi0}), and the kinetic energy of the
excitations
are equally important.
This is taken into account in our derivation of Eqs.~(\ref{f+})
and
(\ref{f-}), relying on the exact expression for $\Psi_0$.
In principle, the exact equations (\ref{f+}) and (\ref{f-}) can
be used
to obtain a systematic expansion of the excitation wavefunctions
and energies in the
$\zeta$ parameter.

In the case of spherical symmetry
($\omega_{i}=\overline{\omega}=\omega$)
the excitations are characterized by
the orbital angular momentum $l$ and its projection $m$.
The solution of Eq.(\ref{W}) has the form
$W\!=\!x^{l/2}P(x)Y_{lm}(\theta,\phi)$,
where $Y_{lm}$ is a \linebreak
-------------------------------------------------------------------------
\vspace{27mm}

\hspace{-4mm}-----------------------------------------------------------------------

\hspace{-4mm}spherical harmonic, $x\!=\!y^2$, and
the radial function $P(x)$ is governed by a hypergeometric
differential equation
\begin{equation}            \label{hyp}
\!\!x(1\!\!-\!x)\frac{d^2P}{dx^2}\!+\!\left[\!l\!+\!{3 \over
2}\!-
\!\left(\!l\!+\!{5\over 2}\!\right)\!x\right]\!\frac{dP}{dx}\!
\!+\!\left(\!{{\varepsilon^{2}}\over 2}\!
-\!{l\over 2}\right)\!\!P\!=\!0.\!\!
\end{equation}
The solution of Eq.(\ref{hyp}), convergent at $x=0$, is the
hypergeometric
function which converges at $x\rightarrow 1$ only when reduced to
a
polynomial.
This immediately gives the energy spectrum:
\begin{equation}           \label{spectrum}
E_{nl}=\hbar\omega\varepsilon_{nl}=\hbar\omega
(2n^{2}+2nl+3n+l)^{1/2},
\end{equation}
where $n$ is a positive integer.
The solutions of Eq.(\ref{hyp}) are classical Jacobi polynomials
$P_{n}^{(l+1/2,0)}(1-2y^{2})$ and, with the normalization
conditions
(\ref{norm}) and (\ref{C}), we obtain
\begin{eqnarray}
f_{\!+}\!\!&=&\!\!\left[\!\frac{(1\!\!-\!y^2)(\!4n\!+\!2l\!\!+\!3)}
{l_c^3\varepsilon_{nl}\zeta}\!\right]^{\!{1/2}}\!\!\!\!\!\!
y^l\!P_{n}^{(l\!+\!1/2,0)}(1\!\!-\!2y^{2})Y_{\!{lm}}(\theta\!,\!\phi)\!\!
  \label{f+1} \\
f_{\!-}\!\!&=&\!\!\left[\frac{\varepsilon_{nl}\zeta(\!4n\!+\!2l\!+\!3)}
{l_c^3(1-y^2)}\!\right]^{\!{1/2}}\!\!\!\!
y^l\!P_{n}^{(l\!+\!1/2,0)}(1\!\!-\!2y^{2})Y_{\!{lm}}(\theta\!,\!\phi),\!
  \label{f-1}
\end{eqnarray}
where $l_c=(2\mu/M\omega^2)^{1/2}$ is the size of the condensate.
The spectrum (\ref{spectrum}) coincides with that found by
Stringari
\cite{Stringari} from the analysis of the density fluctuations in
the
hydrodynamic approach.

In the non-symmetric case with
$\omega_{1}\ne\omega_{2}\ne\omega_{3}$,
the operator $\hat G$ is invariant under the inversion of any of
the
--------------------------------------------------------------------------
\widetext
\begin{eqnarray}             \label{Beq}
\left[ (1-\rho^2-z^2)\left(\frac{\partial^2}{\partial\rho^2}
+\frac{(2|m|+1)}{\rho}\frac{\partial}{\partial\rho}+
\beta^2\frac{\partial^2}{\partial z^2}\right)\!-\!2
\!\left(\rho\frac{\partial}{\partial\rho}\!+\!\beta^2z\frac{\partial}
{\partial z}
\right)\!+\!2\left(\frac{\Omega_{nm}^2}{\omega_{\rho}^2}\!-\!m\right)\right]
B_{nm}(\rho,z)\!=\!0,
\end{eqnarray}
--------------------------------------------------------------------------
\narrowtext
\hspace{-4mm}three spatial coordinate. Therefore, the polynomials
$W$
determined by Eq.(\ref{W}) can be labeled by the corresponding
parities
${\cal P}_{i}=[\pm,\pm,\pm]$. Another quantum number is the order
$N$ of
the polynomial $W$. For even $N$ the function $W$ contains the
powers of
$y$ equal to $N,N-2,...,0$, and for odd $N$ the powers are equal
to
$N,N-2,...1$.

In fact, there are only two independent parities, since
$\prod_i{\cal P}_i=(-1)^N$.
The first few eigenstates can easily be found.
For $N=1$ we have three eigenstates describing the condensate
center of
mass oscillations with the trap frequencies.
Accordingly, there are three eigenfunctions $W\propto y_{i}$ with
parities
${\cal P}_{i}=[-]$. The corresponding eigenfrequencies are
$\Omega=\overline{\omega}\varepsilon=\omega_{i}$.
In the case of $N=2$ we again obtain three eigenstates
corresponding to
the condensate center of mass oscillations. The eigenfunctions
are
$W\propto y_{i}y_{j}$ ($i\ne j$), the parities ${\cal P}_{i}={
\cal P}_{j}=[-]$, and the
eigenfrequencies $\Omega=\sqrt{\omega_{i}^{2}+\omega_{j}^{2}}$.
In addition, there are three eigenstates with $N=2$ and  parities
${\cal P}_{i}=[+]$ for all $i$.
Those correspond to the quadrupole oscillations of the
condensate, the center of mass being at rest.
The eigenfunctions can be written as
$$W\propto 1+\sum_{i=1}^{3}b_i({\bar\omega}/\omega_i)^2y_i^2.$$
There are three different sets of coefficients $b_{i}$
corresponding to the
three eigenfrequencies $\Omega$ determined by the secular
equation
${\rm det}[S]=0$ where $S$ is the $3\times 3$ matrix
$$S_{ij}=1+( 2-\Omega^2/\omega_i^2)
\delta_{i,j}.$$
The coefficients $b_{i}$ are determined by the system
of three linear equations. The first one is
$\sum_{i=1}^{3}b_{i}+(\Omega^{2}/\bar\omega^{2})=0$.
The two other equations can be any of the three linearly
dependent equations
$\sum_{i=1}^{3}S_{ij}b_{i}=0$.

For cylindrically symmetric traps
($\omega_1=\omega_2=\omega_{\rho}$,
$\omega_3=\omega_z$) the projection of the orbital angular
momentum on
the $z$ axis, $m$, is a conserved quantity.
The eigenstates of the
excitations can be labeled by the quantum numbers $N$, $m$, and
the axial
parity ${\cal P}_z$.
The radial parity describing the behavior of the eigenfunctions
with
respect to simultaneous inversion of the two radial coordinates
is $[+]$ for even $m$, and $[-]$ for odd $m$.
The polynomials $W$ can be represented in the form
$W=\rho^{|m|}B_{nm}(\rho,z)\exp{(im\phi)}$,
where $\rho$ and $z$ are the dimensionless axial and radial
coordinates,
and $B_{nm}$ polynomials of power $n=N-m$. Each term of the
polynomial
$B_{nm}$ has the form $z^{n_z}\rho^{n_{\rho}}$, where $n_z$ and
$n_{\rho}$
are positive integers.
The sum $n_z+n_{\rho}$ takes the values $0, 2,...n$ for even $n$,
and
$1, 3...n$ for odd $n$.
The integer $n_{\rho}$ is even, $n_z$ being even for even $n$
(${\cal P}_z=[+]$), and odd for odd $n$ (${\cal P}_z=[-]$).
The polynomials $B_{nm}(\rho,z)$ and eigenenergies
$E_{nm}=\hbar\Omega_{nm}$
can be found from the equation
\linebreak
\vspace{13mm}

\hspace{-4mm}which follows directly from Eq.(\ref{W}).
The quantity $\beta=\omega_z/\omega_{\rho}$ is the ratio of the
axial to
radial frequency.

The number of eigenmodes $k$ at given $m$ and $n$ is determined
by the
quantum number $n$. Generally speaking, we obtain $k=(n+2)/2$ for
even $n$,
and $k=(n+1)/2$ for odd $n$.
The eigenenergies can be found from the $k$-th order secular
equation ${\rm det}[S]=0$, where $S$ is the three-diagonal
$k\times k$ matrix
\begin{eqnarray}         \label{Sij}
&&S_{ij}\!\!=\!\left(\!\frac{\Omega_{nm}^2}{\omega_{\rho}^2}\!-\!|m|\right)
\!\delta_{ij}\!-\!2(k\!-\!j)(k\!-\!i\!+\!|m|\!+\!1\!)
(\delta_{ij}\!+\!\delta_{i-1,j})\! \nonumber\\
&&-\!\beta^2[(i\!-\!1\!+\!q)(2i\!-\!1)\delta_{ij}\!+\!i(2i\!-\!1\!+\!2q)
\delta_{i+1,j}],
\end{eqnarray}
and $i,j=1,2...k$. The coefficient $q=0$ for even $n$, and $q=1$
for odd
$n$.

For $n=0$ ($|m|\geq 1$) we have purely radial oscillations, with
$B_{0m}={\rm const}$ and $\Omega_{0m}=\sqrt{|m|}\omega_{\rho}$.
The case $n=1$ corresponds to the radial oscillations, in
combination
with the axial oscillations of the center of mass of the
condensate.
Here we have $B_{1m}\propto z$ and
$\Omega_{1m}=\sqrt{|m|\omega_{\rho}^2+
\omega_z^2}$.
In both cases the coupling between the radial and axial
motion is absent, and the condensate frequencies $\Omega_{0m}$
and
$\Omega_{1m}$ are the same
as those for a classical gas in the hydrodynamic regime (see
\cite
{Kagan97,Griffin}).

For $n\geq 2$ the coupling between the radial and axial degrees
of freedom
becomes important, and the condensate oscillation frequencies
will be
different from the frequencies of a classical hydrodynamic gas.
If $n=2$ there are two coupled shape oscillations of the
condensate
with frequencies $\Omega_{2m}^{\pm}$ given by
\begin{equation}            \label{omegapm}
\!\!\frac{\Omega_{2m}^{\pm}}{\omega_{\rho}}\!\!=\!\!\left[\!2|m|\!\!+\!\!2\!
+\!\frac{3}{2}\beta^2\!\!\pm\!
\sqrt{\!(|m|\!\!+\!\!2\!-\!\frac{3}{2}\beta^2)^{2}\!\!+\!2\beta^2\!
(|m|\!\!+\!\!1\!)}\!\right]^{\!{\!1\!/\!2}}\!\!\!\!\!\!.\!\!
\end{equation}
In the simplest case of $m=0$ Eq.(\ref{omegapm}) gives the
frequencies
of the quadrupole shape oscillations of the condensate.
The frequencies $\Omega_{20}^{\pm}$ and the frequency of
quadrupole radial
oscillations $\Omega_{02}$ were found in the hydrodynamic
approach in
\cite{Stringari}. They were also obtained in \cite{Castin96} by
considering
the condensate evolution under a weak modulation of the trap
frequencies,
in \cite{Kagan97} from the Hamiltonian of the scaling dynamics,
and in
\cite{Zoller} on the basis of variational approach.
The frequencies $\Omega_{02}$ and $\Omega_{20}^{-}$ were measured
in the
JILA experiment \cite{jila} for $\beta=\sqrt{8}$ and calculated
numerically
for this trapping geometry in \cite{Burnett}. The frequencies
$\Omega_{20}^{\pm}$ were found in the MIT experiment \cite{mit}
for
$\beta=0.08$.

For $n=3$ we have two coupled shape oscillations which are now
also coupled to the oscillations of the center of mass of the
condensate.
In this case we obtain
\begin{equation}        \label{omega3}
\!\!\frac{\Omega_{3m}^{\pm}}{\omega_{\rho}}\!\!=\!\!\left[\!2|m|\!\!+\!\!2
\!+\!\frac{7}{2}\beta^2\!\!\pm\!
\sqrt{\!(|m|\!\!+\!\!2\!-\!\frac{5}{2}\beta^2)^{2}\!\!+\!6\beta^2\!
(|m|\!\!+\!\!1\!)}\!\right]^{\!{\!1\!/\!2}}\!\!\!\!\!\!.\!\!
\end{equation}
Interestingly, for certain values of the projection of the
orbital
angular momentum,
$m$, we find an accidental degeneracy in the spectrum of
excitations.
The simplest example concerns the frequencies $\Omega_{1m}=
\omega_{\rho}\sqrt{|m|+\beta^2}$ and $\Omega_{2m}$.
As follows from Eq.(\ref{omegapm}), $\Omega_{2m}^{-}=\Omega_{1m}$
for
the projection $m$ satisfying the condition
\begin{equation}     \label{c}
\beta^2=|m|+3
\end{equation}
and, accordingly, integer $\beta^2\geq 3$.
For $\beta=\sqrt{3}$ we have $m=0$, i.e., the frequency of
quadrupole
shape oscillations $\Omega_{20}^{-}$ coincides with the frequency
of
axial oscillations $\Omega_{10}$.
In the JILA trapping geometry, where $\beta=\sqrt{8}$,
Eq.(\ref{c}) gives
$|m|=5$.

Although the condensate frequencies $\Omega_{2m}$ do not
significantly differ from those for a classical gas in the
hydrodynamic regime (for $m=0$ see \cite{Kagan97}), the
accidental
degeneracy determined by Eq.(\ref{c}) is characteristic only for
the
condensate.
The presence of the accidental degeneracy can strongly influence
the
picture of the condensate oscillations.
The coupling between the degenerate modes is provided by the
interaction
terms in the Hamiltonian (\ref{H}), proportional to
$\hat\Psi^{\prime 3}$
and $\hat\Psi^{\prime 4}$ and omitted above in the derivation of
the
Bogolyubov-De Gennes equations.
Therefore, driving only one of the degenerate
modes, it is feasible to expect the appearance of oscillations
representing a superposition
of the two modes. This phenomenon ensures a clear distinction
between the condensate oscillations and the oscillations of a
classical
hydrodynamic gas and, hence, can be a signature of
BEC for the gas in the hydrodynamic regime.

E.L.S. and G.V.S. acknowledge the support from the Dutch
Foundation
FOM, from NWO (project 047-003.036), from INTAS, and from the
Russian
Foundation for Basic Studies.
P.\"O, I.T. and M.W. acknowledge hospitality in the group of J.
Mlynek.
M.W. acknowledges financial support by the Deutsche
Forschungsgemeinschaft. I. T. wishes to acknowledge the
fellowship from
the Alexander von Humboldt Foundation.


\begin{references}

\bibitem{Cor95}  M.H. Anderson, J.R. Ensher, M.R. Matthews, C.E.
Wieman,
and E.A. Cornell, Science, {\bf 269}, 198 (1995).

\bibitem{Hul95}  C.C. Bradley, C.A. Sackett, J.J. Tolett, and
R.G. Hulet,
Phys. Rev. Lett., {\bf 75}, 1687 (1995).

\bibitem{Ket95} K.B. Davis, M.-O. Mewes, M.R. Andrews, N.J. van
Druten, D.S. Durfee, D.M. Kurn, and W. Ketterle, Phys. Rev.
Lett.,
{\bf 75}, 3969 (1995).

\bibitem{jila} D.S. Jin, J.R. Ensher, M.R. Matthews, C.E. Wieman,
and
E.A. Cornell, Phys. Rev. Lett., {\bf 77}, 420 (1996).

\bibitem{Mew96} M.-O. Mewes, M.R. Andrews, N.J. van Drutten, D.M.
Kurn, D.S. Durfee, and W. Ketterle, Phys. Rev. Lett., {\bf 77},
416
(1996).

\bibitem{And96} M.R. Andrews, M.-O. Mewes, N.J. van Drutten, D.M.
Kurn,
D.S. Durfee, C.G. Townsend, and W. Ketterle, Science, {\bf 273},
84
(1996).

\bibitem{mit} M.-O. Mewes, M.R. Anderson, N.J. van Drutten, D.M.
Kurn,
D.S. Durfee, C.G. Townsend, and W. Ketterle, Phys. Rev. Lett.,
{\bf
77}, 988 (1996).

\bibitem{Jin96} D.S. Jin, M.R. Matthews, J.R. Ensher, C.E.
Wieman, and
E.A. Cornell, Preprint, 1996.

\bibitem{Stringari} S. Stringari, Phys. Rev. Lett., {\bf 77},
2360
(1996).

\bibitem{Burnett} M. Edwards, P.A. Ruprecht, K. Burnett, R.J.
Dodd,
and C.W. Clark, Phys. Rev. Lett., {\bf 77}, 1671 (1996).


\bibitem{Singh} K.G. Singh and D.S. Rokhsar, Phys. Rev. Lett.,
{\bf
77}, 1667 (1996).

\bibitem{Jav} J. Javanainen, Phys. Rev. A, {\bf 54} 3722 (1996)

\bibitem{Zoller} V.M. Perez-Garcia, H. Michinel, J.I. Cirac, M.
Lewenstein, and P. Zoller, Phys. Rev. Lett., {\bf
77}, 5320 (1996).

\bibitem{You} L. You, W. Hoston, and M. Lewenstein, Phys. Rev. A
{\bf 55},
1581 (1997)

\bibitem{Castin96} Y. Castin and R. Dum, Phys. Rev. Lett., {\bf
77}, 5315
(1996)

\bibitem{Kagan97} Yu. Kagan, E.L. Surkov, and G.V. Shlyapnikov,
Phys.
Rev. A, {\bf 55}, R18 (1997).

\bibitem{Griffin} A. Griffin, W.-C. Wu, and S. Stringari, Phys.
Rev. Lett.,
{\bf 78}, 1838 (1997).

\bibitem{LL} E.M. Lifshitz and L.P. Pitaevskii, {\it Statistical
Physics,
Part 2} (Pergamon Press, Oxford, 1980).

\bibitem{Gennes} P.R. de Gennes, {\it Superconductivity of Metals
and
Alloys}, (Benjamin, New York, 1966).

\bibitem{phase} M. Lewenstein and L. You, Phys. Rev. Lett., {\bf
77}, 3489
(1996).

\bibitem{Silvera}  V.V. Goldman, I.F. Silvera, and A.J. Leggett,
Phys.
Rev. B, {\bf 24}, 2870 (1981).

\bibitem{Huse}  D.A. Huse and E.D. Siggia, J. Low Temp. Phys.,
{\bf 46},
137 (1982).

\end{references}
\end{document}